\documentclass[12pt]{article}

\begin{document}

\vspace*{2.0cm}

\begin{center}
{\large {\bf Hydrodynamical study of neutrino-driven wind 
as an r-process site\\ } }

\vspace*{1.0cm}
K. Sumiyoshi$^{a,}$\footnote{e-mail: sumi@postman.riken.go.jp},
H. Suzuki$^{b}$,
K. Otsuki$^{c}$,
M. Terasawa$^{a,c,d}$,
and S. Yamada$^{d}$ \\
\vspace*{0.5cm}
      $^{a}$The Institute of Physical and Chemical Research (RIKEN), \\
            Hirosawa, Wako, Saitama 351-0198, Japan \\
      $^{b}$High Energy Accelerator Research Organization (KEK), \\
            Oho, Tsukuba, Ibaraki 305-0801, Japan \\
      $^{c}$National Astronomical Observatory (NAO), \\
            Oosawa, Mitaka, Tokyo 181-8588, Japan \\
      $^{d}$Department of Physics, School of Science, The University of Tokyo, \\
            Hongo, Bunkyo, Tokyo 113-0033, Japan \\
%      $^{c}$Research Center for Nuclear Physics(RCNP), Osaka University, \\
%            Mihogaoka, Ibaraki, Osaka 567-0047, Japan \\
\end{center}

\vspace*{0.5cm}
\newpage
%%%%%%%%%%%%%%%%%%%%%%%%%%%%%%%%%%%%%%%%%%%%%
\begin{abstract}
We study the neutrino-driven wind from the proto-neutron star 
by the general relativistic hydrodynamical simulations.  
We examine the properties of the neutrino-driven 
wind to explore the possibility of the r-process nucleosynthesis.  
The numerical simulations with the neutrino heating and cooling 
processes are performed with the assumption of the constant neutrino 
luminosity by using realistic profiles of 
the proto-neutron star (PNS) as well as simplified models.  
The dependence on the mass of PNS and the neutrino luminosity 
is studied systematically.  
Comparisons with the analytic treatment in the previous studies are 
also done.  
In the cases with the realistic PNS, we found that 
the entropy per baryon and the expansion time scale are 
neither high nor short enough for the r-process within 
the current assumptions.  
On the other hand, 
we found that the expansion time scale obtained by the hydrodynamical 
simulations is systematically shorter than that 
in the analytic solutions due to our 
proper treatment of the equation of state.  
This fact might lead to the increase of the neutron-to-seed ratio, 
which is suitable for the r-process in the neutrino-driven wind.  
Indeed, 
in the case of massive and compact proto-neutron stars with 
high neutrino luminosities, the expansion time scale is 
found short enough in the hydrodynamical simulations and 
the r-process elements up to $A \sim 200$ are produced 
in the r-process network calculation.  
\end{abstract}

%%%%%%%%%%%%%%%%%%%%%%%%%%%%%%%%%%%%%%%%%%%%%
\newpage
\section{Introduction}

Finding the origin of heavy elements has been an attractive 
and longstanding issue in astrophysics.  
Among various nucleosynthesis processes, the rapid neutron 
capture process (r-process) is believed to be essential to 
create many of heavy elements (Burbidge et al. 1957).  
Since the r-process nucleosynthesis requires enough neutron 
source to create heavy elements up to $A \sim 200$, 
how and where, in the Universe, the condition for 
the r-process is realized has been a central issue.  
The recent observations of r-process elements in metal poor 
stars (Sneden et al. 1996) further 
strengthen the motivation to identify the r-process site(s) 
in the Universe.  

Among various proposed sites, 
the type II supernovae have been focused as the most 
plausible site and many studies have been made to find out 
where exactly the r-process occurs 
during the event of supernova explosions 
(Hillebrandt 1978; Cowan, Thielemann \& Truran 1991).  
The neutrino-driven wind from a proto-neutron star, just 
born in the supernova explosion, has been suggested as a promising 
site (Meyer et al. 1992; Woosley et al. 1994).  
The surface material of the proto-neutron star is 
heated by the supernova neutrinos and a portion 
of the material is ejected as a hot bubble having 
a high entropy per baryon ($\sim 400 k_{B}$).  
Due to this high entropy, the neutron-to-seed ratio 
after the charged particle reactions freeze out 
($\alpha$-rich freeze-out) becomes high enough 
to lead to the r-process nucleosynthesis.  

The dynamics of the neutrino-driven wind and its outcome 
of the nucleosynthesis have been studied since then 
(Qian \& Woosley 1996 (here after QW); 
Hoffman, Woosley \& Qian 1997; Otsuki 1999, Otsuki et al. 1999).  
Qian and Woosley have investigated the dynamics of the 
neutrino-driven wind both by analytic treatments and 
numerical simulations.  
They have shown that the entropy per baryon in the wind 
turns out too low by a factor of $2-3$ for the r-process.  
Witti et al. (1994) and Takahashi et al. (1994) have 
performed the numerical simulations of the neutrino-driven 
wind and the r-process adopting an initial 
configuration provided by Wilson.  
The entropy per baryon in their simulation falls again 
short for the r-process.  
They had to introduce artificially 
an extra factor to scale down the densities along the 
trajectory to get high entropies for successful r-process.  
These studies, which are mostly done non-relativistically, 
have cast questions on the scenario of 
high entropy bubble for the r-process.  

Cardall and Fuller (1997) have further studied 
the general relativistic dynamics of the neutrino-driven wind with 
the analytic treatment.  
They have shown that the general relativistic effects 
increase the entropy and shorten the expansion time scale 
and will be favorable for the r-process.  
Furthermore, Otsuki et al. (1999) have shown with 
extensive parametric studies that 
the r-process is actually possible in the neutrino-driven wind 
with a short expansion time scale even if the entropy is 
not as high as $\sim 400 k_{B}$.  
They have demonstrated by the network calculation 
that the r-process up to $A \sim 200$ 
can take place for massive and compact neutron 
stars having high neutrino luminosities.  
These general relativistic studies revive some interests 
in the neutrino-driven wind as an r-process site 
and suggest that detailed studies are necessary with general relativity 
for realistic models of proto-neutron stars and supernova neutrinos.  

General relativistic studies have been done so far 
only analytically for stationary neutrino-driven winds 
with given mass and radius of the proto-neutron stars.  
To remove these restrictions, we have performed 
general relativistic hydrodynamical simulations of 
the neutrino-driven wind.  
We have adopted the general relativistic, 
implicit and Lagrangian hydro code, which were developed for the supernova 
simulations (Yamada 1997), and have implemented necessary neutrino 
processes.  
We aim to perform simulations of the time evolution 
of the neutrino-driven wind adopting the results 
of proto-neutron star cooling simulations 
(Suzuki 1994; Sumiyoshi, Suzuki \& Toki 1995).  
The final goal is to find out whether the r-process 
takes place in a realistic situation under the general 
relativistic hydrodynamics and to answer whether the 
high entropy bubble scenario or the rapid expansion 
scenario or any other way is realized to create 
the r-process elements.  

In the current paper, we report the first step of this 
line of research.  
We adopt the realistic profiles of the proto-neutron stars, 
but we take rather simple assumptions about the neutrino 
luminosities and spectra 
in order to compare with the previous analytic studies.  
We remark that the relativistic EOS table, which has 
been completed for supernova simulations 
(Shen et al. 1998; Shen et al. 1998), 
enabled us to construct the initial configurations 
of the wind consistently 
with the proto-neutron star cooling simulations 
using the same physical EOS table.  
We examine systematically the dependence on the 
mass of proto-neutron stars and the neutrino 
luminosities.  

The paper is arranged as follows.  
In section 2, 
after the brief description of the neutrino-driven wind, 
we give explanations of our numerical treatment.  
In section 3, 
we show the numerical results of simulations 
for the realistic profiles of the proto-neutron stars 
($\S$ 3.1) and the simplified models 
with given mass and radius ($\S$ 3.2).  
In section 4, 
we discuss the r-process condition derived from the numerical 
simulations ($\S$ 4.1) and 
describe the r-process network calculation 
using the trajectory of our simulation ($\S$ 4.2).  
We demonstrate the case of successful r-process 
nucleosynthesis from our results.  
The summary will be given in section 5.  

%%%%%%%%%%%%%%%%%%%%%%%%%%%%%%%%%%%%%%%%%%%%%
\section{Hydrodynamical simulation}

The neutrino-driven wind is the mass ejection from the 
surface of the proto-neutron star due to neutrino heating 
(Duncan, Shapiro \& Wasserman 1986).  
Since the proto-neutron star, which is formed in the 
collapse-driven supernova explosion, emits a plenty of 
neutrinos, a small portion of the surface material might be 
heated up and ejected by gaining the energy to escape 
the gravitational potential of the compact object.  
Ejected matter is expanded 
and then cools down to the temperature regime for the nucleosynthesis.  

The main interest here is the thermodynamical history 
of ejecta determined by hydrodynamics during 
the time of the nucleosynthesis.  
If it is favorable 
to make the neutron-to-seed ratio high enough 
during the charged particle reactions, the r-process 
may take place afterwards.  
The key quantities are the electron 
fraction, $Y_{e}$, the entropy per baryon, $S$, and 
the expansion time scale, $\tau_{dyn}$.  
A low electron fraction, a high entropy per baryon 
and a short expansion time scale 
are known to be favorable for the r-process 
(Meyer \& Brown 1997).  
We investigate those conditions by performing the hydrodynamical 
simulations for the surface layers 
of proto-neutron stars.  

\subsection{Numerical code}

We employ the implicit numerical code for the general relativistic 
and spherically symmetric hydrodynamics (Yamada 1997).  
The general relativity is essential here since it 
is known to influence the properties 
of neutrino-driven wind and may lead to a better 
condition for 
the r-process (Cardall \& Fuller, 1997; Otsuki et al. 1999).  
% entropy etc --> intro?
The implicit time differencing is advantageous 
to follow 
the hydrodynamics for a long time compared with 
the sound crossing time of the numerical mesh.  
The hydro code uses a Lagrangian mesh, which is 
suitable to follow the thermal history for the nucleosynthesis.  
We employ the baryon mass meshes with equal spacing.  
The grid size ranges typically from $10^{-8}M_{\odot}$ 
to $10^{-6}M_{\odot}$, depending mainly on the luminosity, 
so as to have enough resolutions.  

The heating and cooling processes due to neutrinos are 
added on top of the hydro code.  
The optically thin-limit is assumed for neutrinos since 
the surface region of interest has low densities and 
is transparent to neutrinos.  
Although the Boltzmann solver of neutrino transfer is 
already implemented in the numerical code 
(Yamada, Janka \& Suzuki 1999), we do not solve the 
Boltzmann equation, but instead we set 
the neutrino distribution function 
at each Lagrangian mesh point as described below.  
% relation of nn and Li
% angle mesh, energy mesh

We treat the following neutrino reactions as sources of 
heating and cooling, 
\begin{eqnarray}
\nu_{e} + n \leftrightarrow e^{-} + p ,\\
\bar{\nu}_{e} + p \leftrightarrow e^{+} + n ,\\
\nu_{i} + e^{-} \leftrightarrow \nu_{i} + e^{-} ,\\
\nu_{i} + e^{+} \leftrightarrow \nu_{i} + e^{+} ,\\
\bar{\nu}_{i} + e^{-} \leftrightarrow \bar{\nu}_{i} + e^{-} ,\\
\bar{\nu}_{i} + e^{+} \leftrightarrow \bar{\nu}_{i} + e^{+} ,\\
\nu_{i} + \bar{\nu}_{i} \leftrightarrow e^{-} + e^{+} ,
\end{eqnarray}
where the index {\it i} of $\nu_{i}$ stands for neutrino 
flavors ($i=e, \mu, \tau$).  
The evaluation of the reaction rates 
(1) and (2) follows the standard 
procedure as used in supernova calculations 
(Yamada, Janka \& Suzuki 1999).  
The heating and cooling rates are calculated by the 
energy integrals using the distribution functions of 
neutrinos and electrons in the Boltzmann solver.  
The Pauli blocking effects for electrons and positrons, 
the neutron-proton mass difference are thus properly taken 
into account.  
% mass difference?
This is in contrast to the approximate treatment in QW 
where the radiation dominated situation is assumed.  
The other heating and cooling rates due to the pair process 
and the electron scattering are evaluated by 
Eqs. (12), (13) and (14) of QW.  
% equation expression
The evolution of the electron fraction, $Y_{e}$, is solved 
together with hydrodynamics using the collision terms 
for the reactions (1) and (2) of the Boltzmann solver 
(Yamada 1997).  
% Ye^{eq}

As for the equation of state (EOS) of dense matter, 
we adopt the table of the relativistic EOS, which 
is recently derived for supernova simulations 
(Shen et al. 1998, Shen et al. 1998) in 
the relativistic nuclear many body framework.  
It reproduces the 
nuclear matter saturation and the properties of 
stable and unstable nuclei in the nuclear chart
(Sugahara \& Toki 1994; Sumiyoshi, Kuwabara \& Toki 1995).  
The table covers the wide range of density 
($10^{5.1} \sim 10^{15.4}$ g/cm$^{3}$), 
electron fraction 
($0.0 \sim 0.56$), 
and temperature 
($0 \sim 100$ MeV), 
which is required for supernova simulations.  
The electron/positron and photon contributions 
as non-interacting particles 
are added to the nuclear contribution of the EOS.  
The arbitrary degeneracy of electrons and 
disappearance of positrons at low temperatures 
are properly treated.  

We extend the EOS table toward lower densities 
below $10^{5}$ g/cm$^{3}$ in the current study.  
Since this low density regime appears in the simulations 
at temperature below 0.5 MeV, 
we assume the mixture of neutrons, protons and 
$\alpha$-particles in nuclear statistical equilibrium.  
This is a good approximation at the time of $\alpha$-rich 
freeze-out because only a slight amount of nuclei is synthesized.  
To determine the composition precisely in this 
temperature regime, one has to solve the nuclear reaction network 
with the hydrodynamics at the same time.  
This is a formidable task beyond the current scope of study.  
We note, however, that the evolution of the electron 
fraction is properly coupled to hydrodynamics.  

As for the neutrino spectra, we take rather simple 
assumptions in the current study.  
The neutrino distribution, $f_{\nu_{i}}(R_{\nu_{i}})$, 
at the neutrinosphere is assumed 
to be monochromatic in energy and isotropic in angle as follows 
\begin{equation}
f_{\nu_{i}}(R_{\nu_{i}}) = 
f_{\nu_{i0}}\delta(E_{\nu_{i}}-\langle E_{\nu_{i}} \rangle).  
%\delta (\mu - 1), 
\end{equation}
The coefficient, $f_{\nu_{i0}}$, is determined by 
the neutrino luminosity, $L_{\nu_{i}}$, and the radius 
of the neutrinosphere, $R_{\nu_{i}}$, as
\begin{equation}
f_{\nu_{i0}}=
\frac{2 \pi L_{\nu_{i}}}{\langle E_{\nu_{i}}\rangle ^{3}R_{\nu_{i}}^{2}},
\end{equation}
where the average energy, $\langle E_{\nu_{i}}\rangle $, $L_{\nu_{i}}$ and 
$R_{\nu_{i}}$ are given as model parameters.  
The number density of neutrinos, $n_{\nu_{i}}$, at radius, 
$r$, is given by
\begin{eqnarray}
n_{\nu_{i}} & = & \frac{1 - x}{4 \pi^{2}} \langle E_{\nu_{i}}\rangle ^{2}f_{\nu_{i0}}\\
& = & \frac{1-x}{2 \pi}\frac{L_{\nu_{i}}}{\langle E_{\nu_{i}}\rangle R_{\nu_{i}}^{2}},
\end{eqnarray}
where $x$ is defined as 
\begin{equation}
x = \left( 1 - \frac{R_{\nu_{i}}^{2}}{r^{2}} \right)^{\frac{1}{2}},
\end{equation}
to take into account the solid angle subtended by the 
neutrinosphere (QW).  
Using this expression of the number density, we set the neutrino 
distribution at radius, $r$, as 
\begin{equation}
f_{\nu_{i}}(r) = f_{\nu_{ir}}\delta(E_{\nu_{i}}-\langle E_{\nu_{i}}\rangle ),
\end{equation}
where the coefficient, $f_{\nu_{ir}}$, is given by 
\begin{equation}
f_{\nu_{ir}}=\frac{2 \pi (1-x) L_{\nu_{i}}}{\langle E_{\nu_{i}}\rangle ^{3}R_{\nu_{i}}^{2}}.
\end{equation}
The positions of the neutrinospheres are assumed 
to be common to all neutrino species in the current study.  
The general relativistic effects 
such as the red-shift and the ray bending 
are not taken into account here.
The above approximation for neutrino spectra is just intended 
to compare with the previous analytic studies.  
More realistic neutrino spectra will be incorporated 
in the forthcoming paper.  

As for the inner boundary condition, the radius, 
the gravitational mass and the baryon mass are given.  
Thermodynamical quantities are set to be the same 
as those at the innermost mesh point.  
We adopt the reflecting condition for the velocity.  
As for the outer boundary, we impose a constant pressure.  
The thermodynamical quantities are again assumed to be 
the same as those at the neighboring mesh point inside.  
% dependence on outer boundary condition
% 
%Things checked

\subsection{Initial models}

The initial models are constructed based on the hydrostatic 
configuration of the proto-neutron stars.  
We take a surface layer of the proto-neutron 
star and remap it to the hydro code as an initial configuration.  
We construct initial models in two ways.  
One is based on the numerical results of the 
proto-neutron star cooling 
(Sumiyoshi, Suzuki \& Toki 1995) as realistic models.  
In the second case, we 
choose the neutron star mass and radius at 
the inner boundary arbitrarily.  

As for the former case, 
the numerical simulations have been performed by solving the 
quasi-static evolution of the proto-neutron star 
with neutrino transport using the multi-energy-group 
flux-limited diffusion scheme (Suzuki 1994).  
In these simulations, we have used the relativistic 
EOS table, which is the same as the one used in the current 
study.  
By so doing, mapping of thermodynamical quantities to the hydro code 
is done consistently.  
We choose the cases of the baryon mass of $1.62M_{\odot}$ 
and $2.00M_{\odot}$.  
Snapshots of density, electron fraction and 
temperature at a certain time of the proto-neutron star 
cooling simulations are picked up to be used as initial 
inputs for the neutrino-driven wind simulations.  
We remark that the proto-neutron star cooling simulations 
start from the initial profile provided by Mayle and Wilson, 
which corresponds to the supernova core 
at 0.4 seconds after the core bounce.  
We cut out the surface layer containing a small baryon mass 
(typically from 
$10^{-6}M_{\odot}$ to $10^{-4}M_{\odot}$ depending 
on the models) 
from the proto-neutron star and map it onto the mesh of 
the hydro code.  
As for the latter case, 
we assume the mass and radius 
of proto-neutron star arbitrarily 
as inputs like the cases in QW.  
We solve the Oppenheimer-Volkoff equation to 
obtain the structure of a thin layer 
above the given radius.  
We assume the temperature and the electron fraction 
are constant in the whole layer and 
take $T=3$ MeV and $Y_{e}=0.25$ in the current study.  
For both cases, 
we have confirmed that the initial configuration settles into 
static state within a short time after remapping 
as long as we switch off the neutrino reactions.  

We take as a reference the average energy of neutrinos as 
\begin{eqnarray}
\langle E_{\nu_{e}}\rangle  & = & 10\quad {\rm MeV},\\
\langle E_{\bar{\nu}_{e}}\rangle  & = & 20\quad {\rm MeV},\\
\langle E_{\nu_{\mu}}\rangle  & = & 30\quad {\rm MeV}. 
\end{eqnarray}
Similar values are chosen in the analytic studies (QW, Otsuki et al. 1999).  
The average energies of $\mu$, $\tau$ neutrinos and anti-neutrinos 
are assumed to be identical.  
We give the neutrino luminosities as inputs and 
assume that they are constant during the simulations.  
We also assume that the luminosities are common to all flavors.  
These simple settings for neutrinos are meant 
to compare with previous studies and to clarify the dependence 
on the luminosity and average energy of neutrinos and 
the mass and radius of proto-neutron stars.  
In the forth-coming paper, we will take more realistic 
neutrino spectra from the proto-neutron star simulations 
taking into account energy distribution, flavor-dependence 
and time-dependence.  

% We fix the pressure at the outer most interface as an outer 
% boundary condition.  
% We assume that the material with a certain pressure stays 
% above the proto-neutron star.  
% If one assumes radiation-dominated matter, the pressure is 
% given by 
% \begin{equation}
% p=\frac{11 \pi^{2}}{180}T^{4}.  
% \label{eq:prad}
% \end{equation}
% We choose the pressure of the outer material typically 
% being $1 \times 10^{22} ergs/s$.  
% This pressure corresponds to the temperature around $0.1 MeV$ 
% using Eq. (\ref{eq:prad}).  
% We remark here that we do not assume the radiation-dominated 
% matter in the numerical simulation, but we use the equation 
% of state properly calculated by the Fermi integrals for 
% electrons and positrons.  
% We explore the dependence of the result on this choice 
% of the pressure later.  

% figures of initial model?

%\section{Proto-neutron star cooling}
% PNSC model, Mb=1.62Msolar, 2Msolar, Neutrino setting
% (t=3, 15, 20 sec)
%
% OV model (Mg=2Msolar, 10km), Neutrino setting
% R_{nu}

%%%%%%%%%%%%%%%%%%%%%%%%%%%%%%%%%%%%%%%%%%%%%%%%%%%
\section{Numerical results}

\subsection{Models based on proto-neutron star cooling simulations}

We start with a model based on 
the proto-neutron star cooling simulations 
with the baryon mass of $1.62M_{\odot}$.  
As an initial configuration, we employ the output at $t=3$ sec 
of the proto-neutron star cooling simulation.  
We set the neutrino luminosities $L_{\nu_{i}}=1 \times 10^{51}$ 
ergs/s for each species.  
The total luminosity, therefore, amounts to $L_{\nu}^{tot}=6 \times 10^{51}$ 
ergs/s.  
The outer boundary pressure is set to be $p_{out}=10^{22}$ dyn/cm$^{2}$.  
% which corresponds to $T_{out}=0.1 MeV$.  
We take the radius of the neutrinosphere as $R_{\nu}=16.4$ km.  
% while the radius at the inner boundary is $16.8 km$.  
%nw069p06 case

We display in Fig. 1 the trajectories of mass elements 
during the simulation.  
The positions in radius are shown as a function of time.  
The trajectories of every 5 mass shells are presented here.  
The surface layers are 
heated by neutrinos and escape from the surface of the 
proto-neutron star one by one, forming the neutrino-driven wind.  
Figure 2 depicts the temperatures of the mass elements as a 
function of time.  
The temperature once becomes as high as 3 MeV due to the 
neutrino-heating and cools down to 0.1 MeV during the 
expansion.  
The pressure of the material balances 
with the imposed outer pressure when the 
temperature of the material becomes around 0.1 MeV, 
which roughly corresponds to this pressure.  
Figure 3 shows the densities of the mass elements as 
a function of time.  
The density of the material in the wind decreases due to 
expansion and stays around $10^{4}$ g/cm$^{3}$.  
The entropy per baryon, $S$, becomes high during the evolution 
and reaches up to $87$ (in units of $k_{B}$ hereafter) 
at $T=0.5$ MeV in this model.  After that, the entropy 
remains roughly constant because the heating and cooling 
processes become negligible.  

We define the expansion time scale, $\tau_{dyn}$, 
as the $e$-fold time of temperature at $T=0.5$ MeV 
during the expansion.  
It is found to be 0.15 seconds in this model.  
This definition accords with those in the previous 
papers by Hoffman et al. (1997) and Otsuki et al. (1999) to discuss 
on the $\alpha$-rich freeze-out and the r-process.  
The mass loss rate, $\dot{M}$, which is the amount of the 
ejected mass divided by the mass loss time, is found to 
be $1.5 \times 10^{-5}M_{\odot}$/s.  

Figure 4 displays the time evolution of the electron fraction 
for one of the mass elements.  
The electron fraction is small at the beginning 
since the mass element was originally in the surface layer 
of the proto-neutron star.  
It increases due to neutrino reactions on nucleons during 
the expansion in the wind and reaches 0.46 at 
$T=0.5$ MeV.  
The equilibrium value of the electron fraction is roughly 
determined by the neutrino spectra.  
We will discuss this point in section 4.1.  
% Ye^{eq}, Alpha effect

To explore their dependences, 
we have performed simulations with different neutrino luminosities 
and simulations with a different proto-neutron star mass.  
For the latter case, 
we employ the profile of a more massive proto-neutron star 
at 3 seconds in the cooling simulation.  
Table 1 summarizes the model parameters and some results 
of the numerical simulations.  

Figure 5 shows the dependence of the key quantities 
on the neutrino luminosity.  
For comparisons, we plot also the values obtained with 
the analytic treatment (Otsuki et al. 1999), 
in which the neutron star mass and the radius are 
assumed to be $1.4M_{\odot}$ and 
10 km, respectively.  
The dependence on the neutrino luminosity is found 
qualitatively similar between the numerical simulations 
and the analytic treatment despite the difference of 
initial profile.  
We see, however, quantitative differences in the mass loss rate 
and the entropy per baryon.  
The mass loss rate in the current numerical study turns 
out larger than that in the analytic study while 
the entropy per baryon tends to be lower.  
% This is because the radius of the proto-neutron star 
% is rather large due to the finite temperature inside.  
% dependence on neutrino average energy

\subsection{Models with given mass and radius}

To compare with the analytic treatment in detail, we have 
performed the numerical simulations for the proto-neutron star 
with the mass and radius given by hand.  
This is meant to check the numerical results by the comparison 
with the analytic treatment for simple cases as well as 
to understand the results obtained in the previous 
section better.  

Table 2 summarizes the model parameters and some results 
of the numerical simulations.  
We choose $1.4M_{\odot}$ and $2.0M_{\odot}$ for the mass 
and 10 km for the radius to compare with the results by 
Otsuki et al. (1999), where the general relativistic analytic 
study has been worked out for the same parameters.  

Figure 6 shows the key quantities of the neutrino-driven 
wind as a function of neutrino luminosity
both for the numerical simulations and the analytic treatment.
The general trend of the luminosity dependence is common.  
As for the mass loss rate and the entropy per baryon, 
the results of numerical simulations accord quantitatively 
well with those of analytic treatment.  
On the other hand, the expansion time scale is found 
systematically shorter in the numerical simulations 
than in the analytic study.  
The difference becomes large in some cases, 
which is preferable for r-process.  

We have found that this difference of the expansion time 
scale mainly arises from 
our proper treatment of the equation of state.  
In the analytic studies done so far, 
they used the equation of state 
of the radiation-dominated matter 
\begin{equation}
p=\frac{11 \pi^{2}}{180}T^{4}.  
\label{eq:prad}
\end{equation}
even for temperatures below 0.5 MeV.  
It overestimates the pressure of electrons/positrons 
in the low temperature regimes.  
As a result, the pressure at large radii, 
where the temperature is low, 
is larger in the analytic treatment than 
in our simulations with the same temperature.  
The larger pressure results in a longer expansion time 
scale since it decelerates the wind.  

% figures T, v vs r
We have investigated this issue in detail for 
model c01.  
In the numerical simulations, we impose a constant 
pressure ($1 \times 10^{22}$ dyn/cm$^{3}$ for model c01) 
at the outer boundary while in the analytic 
treatment the temperature is 
assumed to be 0.1 MeV at the radius $10^{4}$ km.  
% In the case of the model c01 with the pressure of the 
% outer material $1 \times 10^{22} ergs/s$, the temperature 
% at the radius $10^{3} km$ is found $0.118 MeV$.  
% In the corresponding model by the analytic treatment 
% in Figure 6, 
% as the outer boundary condition.  
% In this analytic result, the temperature at the radius 
% $10^{3} km$ is $0.115 MeV$, which is slightly lower than 
% the simulation.  
% Even though the temperature is slightly lower, 
% the pressure in the analytic treatment is found larger 
% than the pressure in the simulation.  
Although the temperatures at the radius $10^{3}$ km are 
0.12 MeV in both cases, 
the pressures there are found $2.2 \times 10^{22}$ dyn/cm$^{2}$ 
in the analytic treatment and $1.1 \times 10^{22}$ dyn/cm$^{2}$ 
in the simulation.  
This discrepancy comes from different treatments of 
the equation of state and the smaller pressure 
results in a shorter expansion time in the 
simulation.  
To confirm this interpretation, we recalculate the analytic 
model with the lower temperature (0.09 MeV) which gives the 
the same pressure at the radius $10^{3}$ km as model c01.  
In this case, the profiles of the pressure become closer 
to each other, 
although the pressure in the analytic treatment is still 
higher inside the radius $10^{3}$ km.  
The expansion time scale becomes shorter ($0.10$ sec) 
than the original analytic model ($0.16$ sec) and closer 
to the value ($0.05$ sec) in the simulation.  
Thus it is clear that the proper treatment of the equation 
of state is crucial to obtain the expansion time scale.  
We stress again that the shorter expansion time scale 
is preferable for r-process.  

We have also examined the dependence on the pressure 
imposed at the outer boundary.  
When we take a smaller pressure ($1 \times 10^{20}$ dyn/cm$^{2}$) 
the expansion time scale becomes shorter (0.026 sec) than the 
value (0.05 sec) in the standard case ($1 \times 10^{22}$ 
dyn/cm$^{2}$), 
while the other quantities such as the mass loss rate and 
the entropy per baryon change little.  
The temperature 
becomes as low as 0.04 MeV due to the rapid expansion, 
which roughly corresponds to the pressure imposed 
at the outer boundary.  
The radius and the density become larger and lower, 
correspondingly.  
% lower outer pressure --> r-process preferable
% too low outer pressure --> r-process temperature

Here we comment on the effect of the general relativity.  
As pointed out by Cardall \& Fuller (1997) and Otsuki et al. (1999), 
we found that 
the entropy per baryon is higher and 
the expansion time scale is shorter than the 
corresponding results in the Newtonian treatment 
by Qian \& Woosley (1996).  

% reason of low S for pnsc model --> large radius, 2M-15km

%%%%%%%%%%%%%%%%%%%%%%%%%%%%%%%%%%%%%%%%%%%%%
\section{R-process nucleosynthesis}

\subsection{Conditions for r-process}

Based on the hydrodynamical simulations of 
the neutrino-driven wind, 
we discuss here key quantities 
for the r-process nucleosynthesis, that is, 
the entropy per baryon, expansion time scale 
and electron fraction at the time of the nucleosynthesis.  
Higher entropy per baryon, shorter expansion time scale 
and lower electron fraction (and their combinations) 
are favorable for the r-process nucleosynthesis 
(Hoffman et al. 1997; Meyer and Brown 1997).  

Figure 7 summarizes all the results of the numerical simulations 
listed in Tables 1 and 2 
in the plane of the expansion time scale and the entropy 
per baryon.  
The analytic models are also shown in the same figure.  
It is found that the results for the models based on the 
proto-neutron star cooling simulations are similar 
to the analytic model with $1.4M_{\odot}$.  
Among them, the more 
massive case with $2.00M_{\odot}$ is rather 
preferable for the r-process having higher entropy 
per baryon.  
As for the models with given mass and radius, they have 
higher entropy and shorter expansion time scale 
(upper left direction in the figure) than the models 
based on the proto-neutron star cooling simulations.  
This is mainly because the former has a smaller radius 
and a deeper gravitational potential than the latter 
(for details, see Qian \& Woosley 1996).  
We can see here again the general trend that the more massive 
models are the better for the r-process.  
Using analytic model with general relativity, 
Otsuki et al. (1999) pointed out that the r-process might be 
possible for massive neutron stars with a short expansion 
time scale.  
Our results with the numerical simulations 
show even shorter expansion time scale 
having a higher possibility of the successful r-process.  
Indeed, in the analytic study by Otsuki et al. (1999), 
the r-process is possible only for high 
luminosity cases, while the current study 
relaxes this constraint.  

Another key quantity is the electron fraction.  
If the electron fraction is small enough, the r-process 
is possible even with low entropy per baryon and with long 
expansion time scale.  
The electron fraction in the neutrino-driven wind 
is governed by the neutrino captures on nucleons, 
thereby determined by the relative strength of 
the luminosities and energy spectra 
of the electron-type neutrinos and 
anti-neutrinos.  
In the model with a higher average energy for the electron-type 
anti-neutrinos, 
$\langle E_{\bar{\nu}_{e}}\rangle  = 30$ MeV,
the electron fraction at $T=0.5$ MeV turns out to be 
0.38 which is lower than 
0.46 in model p06 with $\langle E_{\bar{\nu}_{e}}\rangle = 20$ MeV.  

The electron fraction in the neutrino-driven wind 
can be estimated by the average energies of 
neutrinos and anti-neutrinos as 
\begin{equation}
Y_{e}^{eq} = \frac{\langle E_{\nu_{e}}\rangle  + 2 \Delta}
               {\langle E_{\nu_{e}}\rangle  +\langle E_{\bar{\nu}_{e}}\rangle },  
\end{equation}
where $\Delta$ is the neutron-proton mass difference 
(Qian and Woosley 1996).  
Here we assume that 
the charge-changing neutrino capture reactions 
are in equilibrium and that 
the luminosity of neutrinos are the same as 
that of anti-neutrinos.  

The electron fractions obtained in the numerical 
simulations roughly accord with the above estimation.  
We found that the neutrino capture reactions 
come to equilibrium before the temperature 
falls down to 0.5 MeV.  
When the temperature goes below 0.5 MeV, the 
mass fraction of $\alpha$-particle becomes large and 
that of proton becomes negligible.  
In this situation, the electron fraction 
remains almost constant 
with only a slight increase due to 
the anti-neutrino capture on neutrons.  
The neutrino capture on protons has already 
frozen out due to the lack of free protons.  
% check Xp, Xn, Xalpha for p06, p10

\subsection{R-process network calculation}

We have performed a 
nucleosynthesis calculation using a result of our 
hydrodynamical simulations for neutrino-driven winds 
in order to demonstrate that the r-process nucleosynthesis 
is indeed possible for the model.  

We employ model b09 (the mass $2.0M_{\odot}$ and 
the radius 10 km with $L_{\nu}^{tot}=6 \times 10^{52}$ ergs/s) 
which has the shortest expansion time scale among our models.  
We pick up one of mass trajectories in the hydrodynamical 
result and input it to the r-process calculation.  
We start the r-process calculation at the time 
when the temperature becomes $T_{9}\equiv \frac{T}{10^{9} {\rm K}}=9$ 
in the trajectory.  
Figure 8 displays the temperature and density 
trajectories taken from the hydrodynamical simulation.  
The origin of time in the figure is shifted so that 
the temperature becomes $T_{9}=9$ there.  
In the network calculation, 
matter is assumed to be initially composed of 
neutrons and protons 
with the electron fraction $Y_{e}=0.44$, 
the result of the hydrodynamical simulation.  
% b09 figure, number of trajectory
% extrapolation after 0.02 sec

The nuclear reaction network employed in the code covers 
neutron-rich nuclei from the $\beta$ stability line 
to the neutron drip line for $Z=10 \sim 100$.  
It includes also light nuclei which are required to synthesize seed 
elements for the r-process.  
The neutron capture, its reversed reaction, 
$\beta$ decay and $\beta$ delayed emission 
are incorporated for $Z \geq 10$.  
On top of the reactions for the r-process, 
charged particle reactions are included to 
follow the $\alpha$-rich freeze-out.  
% All of charged particle reactions for $Z \geq 10$ 
% and $A \leq 28$ are completely implemented.  
The ($\alpha$, n) reactions up to 
$Z=36$ and ($\alpha$, $\gamma$) reactions 
up to $^{28}$Si are included.  
The $\alpha$ reactions that produce 
$^{9}$Be, $^{12}$C and beyond are also included.  
% The reaction network code is based on the work by 
% Meyer et al. (1992).  
The details of the reaction network code will be 
described elsewhere (Terasawa et al. 1999).  
% references of n-cap, beta-decay etc.
% rates from big-bang code
% neutrino reactions?

The neutron-to-seed ratio at $T_{9}=2.5$ is found to be 120 
in the r-process network calculation.  
The electron fraction at the same temperature is 0.42.  
The rapid expansion during the time of the $\alpha$-rich 
freeze out ($T \sim 0.5$ MeV) leads to 
only a small amount of the seed elements resulting in 
the high neutron-to-seed ratio.  
With this high neutron-to-seed ratio, 
the r-process elements up to $A \sim 200$ are produced 
ensuingly.  
Figure 9 shows the yields of our r-process 
network calculation.  
% final abundance when
It is remarkable that the 3rd peak at $A=195$ as well as 
the 2nd peak at $A=130$ is 
reproduced appropriately.  
It is emphasized that this result is consequence of 
the short expansion time scale and 
accords with the result shown by Otsuki et al. (1999).  
The shorter expansion time scale obtained in the current 
simulation than in the analytic treatment 
further enhances the 3rd peak height 
having a higher neutron-to-seed ratio.  
% so what?

%%%%%%%%%%%%%%%%%%%%%%%%%%%%%%%%%%%%%%%%%%%%%
%\section{Discussion}

%Possibility of r-process
% time dependence

%other profile of pnsc at different time
%initial condition of pnsc

%boundary condition
%outer pressure, shock front constrain?

%stationary?

% shock?

%mass dependence >> obs metal poor >> massive star
%
%%%%%%%%%%%%%%%%%%%%%%%%%%%%%%%%%%%%%%%%%%%%%
\section{Summary}

We study numerically the general relativistic 
hydrodynamics of the neutrino-driven 
wind blown from the proto-neutron stars and 
examine the r-process nucleosynthesis there.  
We focus on the entropy per baryon, electron fraction, 
expansion time scale and mass loss rate 
as key quantities for the r-process.  
By employing the results of the proto-neutron star 
cooling simulations as inputs for the neutrino-driven 
wind simulations, we investigate those quantities 
in realistic situations.  
We also make comparison with the results of 
analytic treatment by assuming masses and radii 
of proto-neutron stars as in the analytic study.  
We explore the dependence on the profile of the 
proto-neutron star as well as neutrino luminosity.  

We find that the entropy per 
baryon and the expansion time scale are neither high 
nor short enough for the r-process in the models based 
on proto-neutron star cooling simulations.  
This is mainly because the 
radius of the proto-neutron star is rather large due 
to the thermal pressure.  
The expansion time scale has a strong dependence 
on the neutrino luminosity and 
it becomes as short as 10 msec for high neutrino 
luminosities, which is close to the value required 
in the rapid expansion scenario by Otsuki et al. (1999).  

On the other hand, the hydrodynamical simulations 
for models with given mass and radius show that 
larger masses and smaller radii are found more 
favorable for the r-process.  
In the case of massive and compact neutron stars with high 
neutrino luminosities, the entropy per baryon 
is high enough and the expansion time scale is 
short enough.  
We demonstrate by the network calculation 
a successful r-process nucleosynthesis 
for a model of our hydrodynamical simulations.  
The 2nd and 3rd abundance peaks of r-process elements and 
their relative height are well reproduced.  

We find that the expansion time scales obtained 
in the simulations are systematically shorter 
than the values in the analytic treatment.  
This is because 
the pressure at low temperature 
is overestimated in the analytic treatment, and therefore 
the expansion time scale becomes longer than 
what one expects with the proper treatment of the 
equation of state as in our study.  
Since the expansion time scale determines the 
neutron-to-seed ratio crucially, even a small 
decrease in the time scale may increase 
the neutron-to-seed ratio substantially making 
ensuing r-process more promising.  

We notice here that the dynamics of the neutrino-driven 
wind is sensitive to the outer boundary condition.  
A decrease of the outer boundary pressure results in 
a shorter expansion time scale.  
On the other hand, the temperature should remain 
around 0.1 MeV for about 1 second so that the r-process 
could occur.  
If the outer boundary pressure is too small, 
the temperature decreases too rapidly for 
the r-process to proceed 
even though the expansion 
time scale is short enough for the high 
neutron-to-seed ratio.  
So far most of the previous studies adopted $T=0.1$ MeV 
at $10^{4}$ km as a boundary condition, however, we 
have to bare in mind this additional uncertainty.  

The fact that the results of hydrodynamical simulations 
based on the proto-neutron star cooling simulation 
turns out not suitable for the r-process does 
not exclude the possibility of the r-process 
in the neutrino-driven wind.  
We have assumed the neutrino luminosity is constant 
in time in the current study.  
This is too simple an assumption.  
In reality, however, 
the neutrino luminosity decreases with time 
scale of 10 seconds and one should not 
forget the time dependence of the luminosity.  
In the simulation of Woosley et al. (1994), the 
r-process takes place in a late stage with a 
high entropy when the neutrino luminosity is 
low and the outer material is already expanding 
due to the high neutrino luminosity in the earlier stage.  
The numerical simulations with the time dependent 
luminosities as well as realistic neutrino energy spectra 
are in progress and will be reported elsewhere.  

%low luminosity
%%%%%%%%%%%%%%%%%%%%%%%%%%%%%%%%%%%%%%%%%%%%%
\section*{Acknowledgment}

We are grateful to S. Wanajo, T. Kajino and I. Tanihata 
for encouraging comments and fruitful discussions 
on the r-process nucleosynthesis.  
We would appreciate K. Oyamatsu, H. Shen and H. Toki 
for the advice on the usage of the relativistic EOS table.  
K. S. would like to express special thanks to H. Shen 
for providing the numerical code for the EOS with 
$\alpha$-particles.  
The numerical simulations have been performed on the 
supercomputer VPP700E/128 at RIKEN and VPP500/80 
at KEK (KEK Supercomputer Projects No.98-35 and No.99-52).  
This work is partially supported by the Grants-in-Aid for the
Center-of-Excellence (COE) Research of the ministry of Education,
Science, Sports and Culture of Japan to RESCEU (No.07CE2002).

%%%%%%%%%%%%%%%%%%%%%%%%%%%%%%%%%%%%%%%%%%%%%
\newpage

\section*{References}

\noindent
Burbidge, E.M., Burbidge, G.R., Fowler, W.A. and Hoyle, F. 1957, 
Rev. Mod. Phys. 29, 547\\
\noindent
Cardall, C.Y. and Fuller, G.M. 1997,
ApJ 486, L111\\
\noindent
Cowan, J.J., Thielemann, F.-K. and Truran, J.W. 1991,
Phys. Rep. 208, 267\\
\noindent
Duncan, R.C., Shapiro, S.L. and Wasserman I. 1986,
ApJ 309, 141\\
% \noindent
% Cowan, J.J., McWilliam, A., Sneden, C. and Burris D.L. 1997,
% ApJ 480, 246.\\
\noindent
Hillebrandt, W. 1978,
Space Sci. Rev. 21, 639\\
\noindent
Hoffman, R.D., Woosley, S.E. and Qian, Y.-Z. 1997,
ApJ 482, 951\\
\noindent
K\"appeler, F., Beer, H. and Wisshak, K. 1989,
Rep. Prog. Phys. 52, 945\\
\noindent
Meyer, B.S., Mathews, G.J., Howard, W.M., Woosley, S.E. and Hoffman, R.D. 1992,
ApJ 399, 656\\
\noindent
Meyer, B.S. and Brown, J.S. 1997,
ApJS 112, 199\\
\noindent
Otsuki, K. 1999, Ph.D Thesis, Osaka University\\
\noindent
Otsuki, K., Tagoshi, H., Kajino, T. and Wanajo, S. 2000, 
ApJ 531, in press\\
\noindent
Qian, Y.-Z. and Woosley, S.E. 1996, 
ApJ 471, 331 (QW)\\
\noindent
Shen, H., Toki, H., Oyamatsu, K. and Sumiyoshi, K. 1998, 
Nucl. Phys. A637, 435\\
\noindent
Shen, H., Toki, H., Oyamatsu, K. and Sumiyoshi, K. 1998, 
Prog. Theor. Phys. 100, 1013\\
\noindent
Sneden, C., McWilliam, A., Preston, G.W., Cowan, J.J., Burris, D.I. 
and Armosky, B.J. 1996,
ApJ 467, 819\\
\noindent
Sumiyoshi, K., Kuwabara, H. and Toki, H. 1995,
Nucl. Phys. A581, 725\\
\noindent
Sumiyoshi, K., Suzuki, H. and Toki, H. 1995,
A\&A 303, 475\\
\noindent
Sugahara, Y. and Toki, H. 1994,
Nucl. Phys. A579, 557\\
\noindent
Suzuki, H. 1994,
Physics and Astrophysics of Neutrinos, 
edited by Fukugita, M. and Suzuki, A., 
(Springer-Verlag, Tokyo, 1994), p763\\
\noindent
Takahashi, K., Witti, J. and Janka, H.-Th. 1994, 
A\&A 286, 857\\
\noindent
Terasawa, M., Sumiyoshi, K., Tanihata, I. and Kajino, T., 
in preparation\\
\noindent
Witti, J., Janka, H.-Th. and Takahashi, K. 1994, 
A\&A 286, 841\\
\noindent
Woosley, S.E., Wilson, J.R., Mathews, G.J., Hoffman, R.D. and Meyer, B.S. 1994,
ApJ 433, 229\\
\noindent
Yamada, S. 1997, 
ApJ 475, 720\\
\noindent
Yamada, S., Janka, H.-Th. and Suzuki, H. 1999, 
A\&A 344, 533\\

%%%%%%%%%%%%%%%%%%%%%%%%%%%%%%%%%%%%%%%%%%%%%%%%%%%%%%%%%%%%%%%%%%%%
\newpage

\section*{Table caption}

\begin{description}

\item[Table 1]
Summary of models based on proto-neutron star cooling simulations.  
$M_{B}$, $M_{G}$ and $R$ are the baryon mass, gravitational mass 
and radius of the proto-neutron star.  For the definition of the 
other entries, see the text.  

\item[Table 2]
Summary of models with given mass and radius.  
$M_{G}$ and $R$ are gravitational mass 
and radius of the proto-neutron star.  For the definition of the 
other entries, see the text.  
\end{description}

%%%%%%%%%%%%%%%%%%%%%%%%%%%%%%%%%%%%%%%%%%%%%%%%%%%%%%%%%%%%%%%%%%%%
\newpage
\begin{center} {\bf Table 1}
\end{center}
\[\begin{tabular}{ccccccccc} \hline
Model & $M_{B}$ [$M_{\odot}$] & $M_{G}$ [$M_{\odot}$] & $R$ [km] & $L_{\nu}^{tot}$ [ergs/s] & $S$ [$k_{B}$] & $\tau_{dyn}$ [sec] & $\dot{M}$ [$M_{\odot}$/s] & $Y_{e}$ \\ \hline
p07 & 1.62 & 1.55 & 17.7 & $3.6 \times 10^{51}$ & 100 & $3.2 \times 10^{-1}$ & $6.1 \times 10^{-6}$ & 0.47 \\
p06 & 1.62 & 1.55 & 17.7 & $6.0 \times 10^{51}$ &  87 & $1.5 \times 10^{-1}$ & $1.5 \times 10^{-5}$ & 0.46 \\
p08 & 1.62 & 1.55 & 17.7 & $1.8 \times 10^{52}$ &  76 & $5.3 \times 10^{-2}$ & $1.0 \times 10^{-4}$ & 0.45 \\
p09 & 1.62 & 1.55 & 17.7 & $6.0 \times 10^{52}$ &  65 & $1.6 \times 10^{-2}$ & $8.6 \times 10^{-4}$ & 0.44 \\
r06 & 2.00 & 1.88 & 16.9 & $3.6 \times 10^{51}$ & 128 & $2.9 \times 10^{-1}$ & $3.4 \times 10^{-6}$ & 0.46 \\
r03 & 2.00 & 1.88 & 16.9 & $6.0 \times 10^{51}$ & 119 & $1.5 \times 10^{-1}$ & $8.3 \times 10^{-6}$ & 0.46 \\
r12 & 2.00 & 1.88 & 16.9 & $1.8 \times 10^{52}$ & 105 & $5.0 \times 10^{-2}$ & $5.6 \times 10^{-5}$ & 0.45 \\
r01 & 2.00 & 1.88 & 16.9 & $6.0 \times 10^{52}$ &  84 & $1.5 \times 10^{-2}$ & $4.7 \times 10^{-4}$ & 0.44 \\ \hline
\end{tabular} \]
% Mg and R are from proto-neutron star model mw88r02 and mw88r03

%\vspace*{2cm}
\newpage

\begin{center} {\bf Table 2}
\end{center}
\[\begin{tabular}{cccccccc} \hline
Model & $M_{G}$ [$M_{\odot}$] & $R$ [km] & $L_{\nu}^{tot}$ [ergs/s] & $S$ [$k_{B}$] & $\tau_{dyn}$ [sec] & $\dot{M}$ [$M_{\odot}$/s] & $Y_{e}$ \\ \hline
c02 & 1.40 & 10.0 & $3.6 \times 10^{51}$ & 142 & $9.7 \times 10^{-2}$ & $2.7 \times 10^{-6}$ & 0.46 \\
c01 & 1.40 & 10.0 & $6.0 \times 10^{51}$ & 131 & $5.0 \times 10^{-2}$ & $6.5 \times 10^{-6}$ & 0.45 \\
c08 & 1.40 & 10.0 & $1.8 \times 10^{52}$ & 121 & $1.2 \times 10^{-2}$ & $4.4 \times 10^{-5}$ & 0.44 \\
c04 & 1.40 & 10.0 & $6.0 \times 10^{52}$ &  95 & $6.2 \times 10^{-3}$ & $4.2 \times 10^{-4}$ & 0.44 \\
b17 & 2.00 & 10.0 & $3.6 \times 10^{51}$ & 263 & $1.0 \times 10^{-1}$ & $8.8 \times 10^{-7}$ & 0.46 \\
b10 & 2.00 & 10.0 & $6.0 \times 10^{51}$ & 239 & $5.3 \times 10^{-2}$ & $2.2 \times 10^{-6}$ & 0.45 \\
b18 & 2.00 & 10.0 & $1.8 \times 10^{52}$ & 196 & $9.6 \times 10^{-3}$ & $1.6 \times 10^{-5}$ & 0.44 \\
b09 & 2.00 & 10.0 & $6.0 \times 10^{52}$ & 165 & $5.1 \times 10^{-3}$ & $1.3 \times 10^{-4}$ & 0.44 \\ \hline
\end{tabular} \]

%%%%%%%%%%%%%%%%%%%%%%%%%%%%%%%%%%%%%%%%%%%%%%%%%%%%%%%%%%%%%%%%%%%%
\newpage

\section*{Figure captions}

\begin{description}

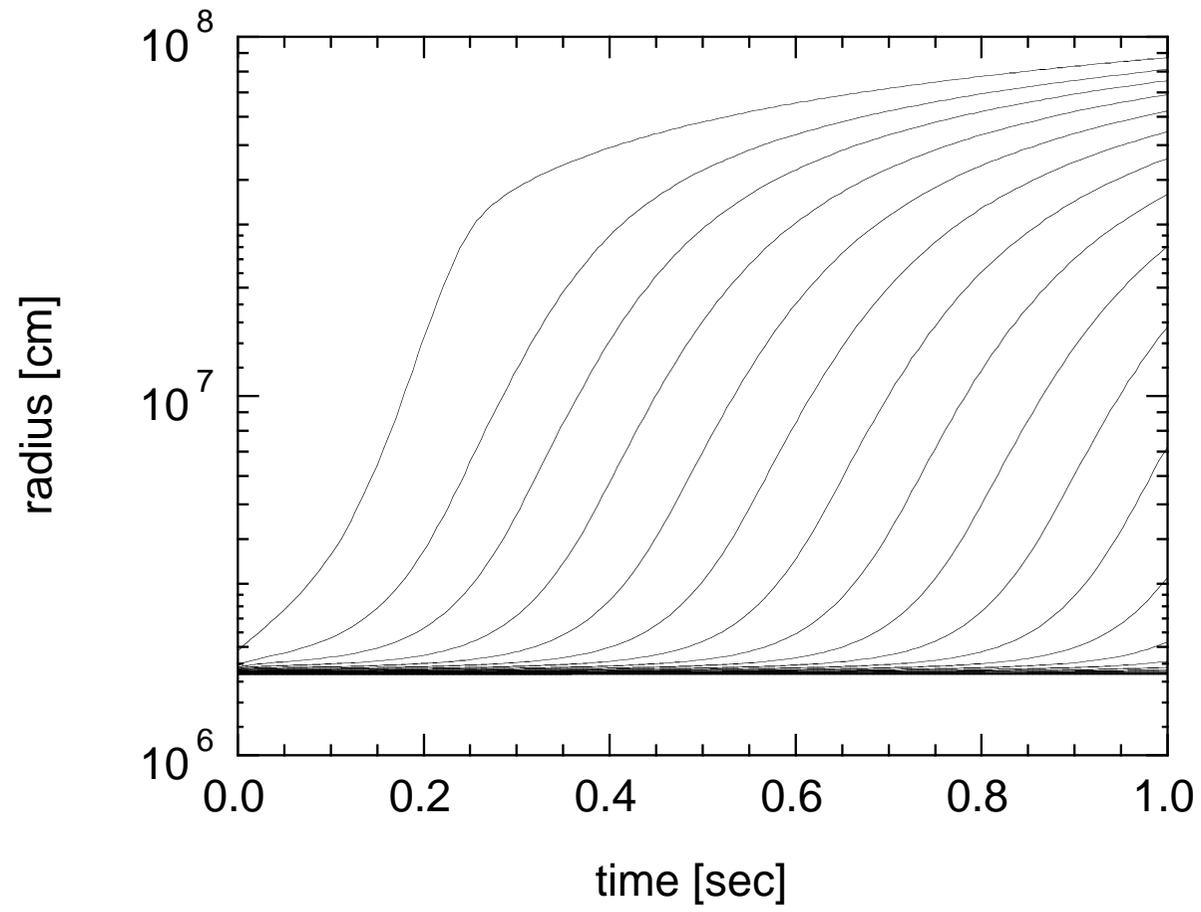
\item[Figure 1]
The trajectories of the mass elements in 
the neutrino-driven wind for model c01 as a function of time.  
Every 5 mass shells are shown here.  

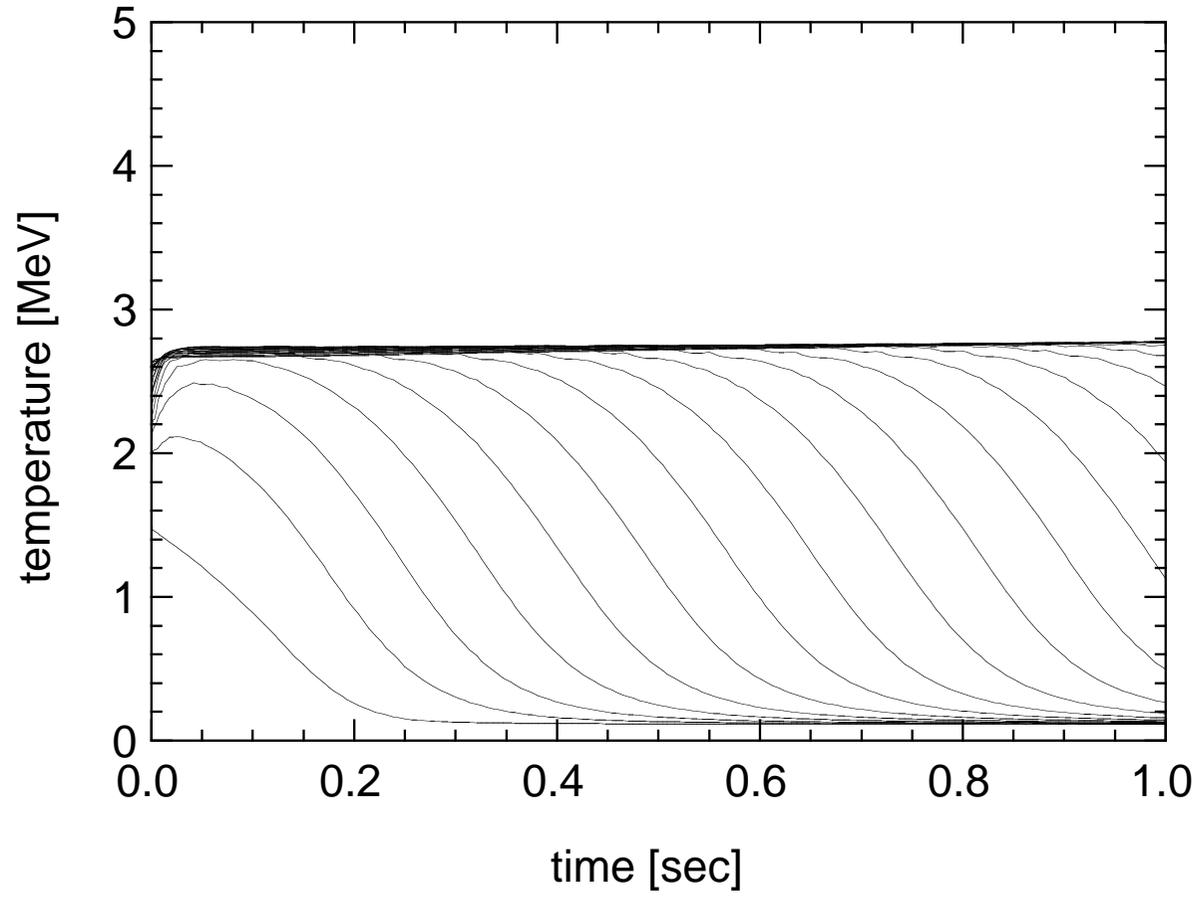
\item[Figure 2]
The temperatures of mass elements for the same model as Fig. 1.  

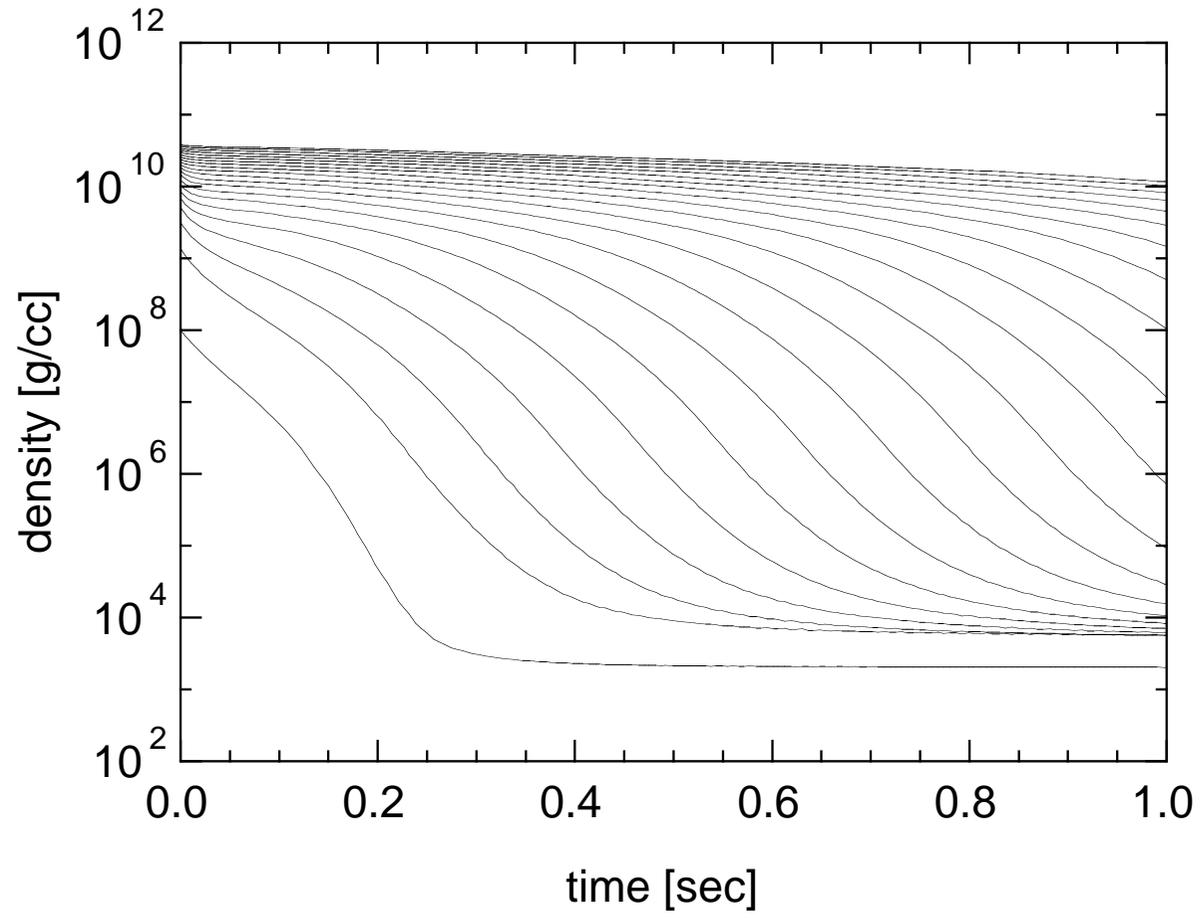
\item[Figure 3]
The densities of mass elements for the same model as Fig. 1.  

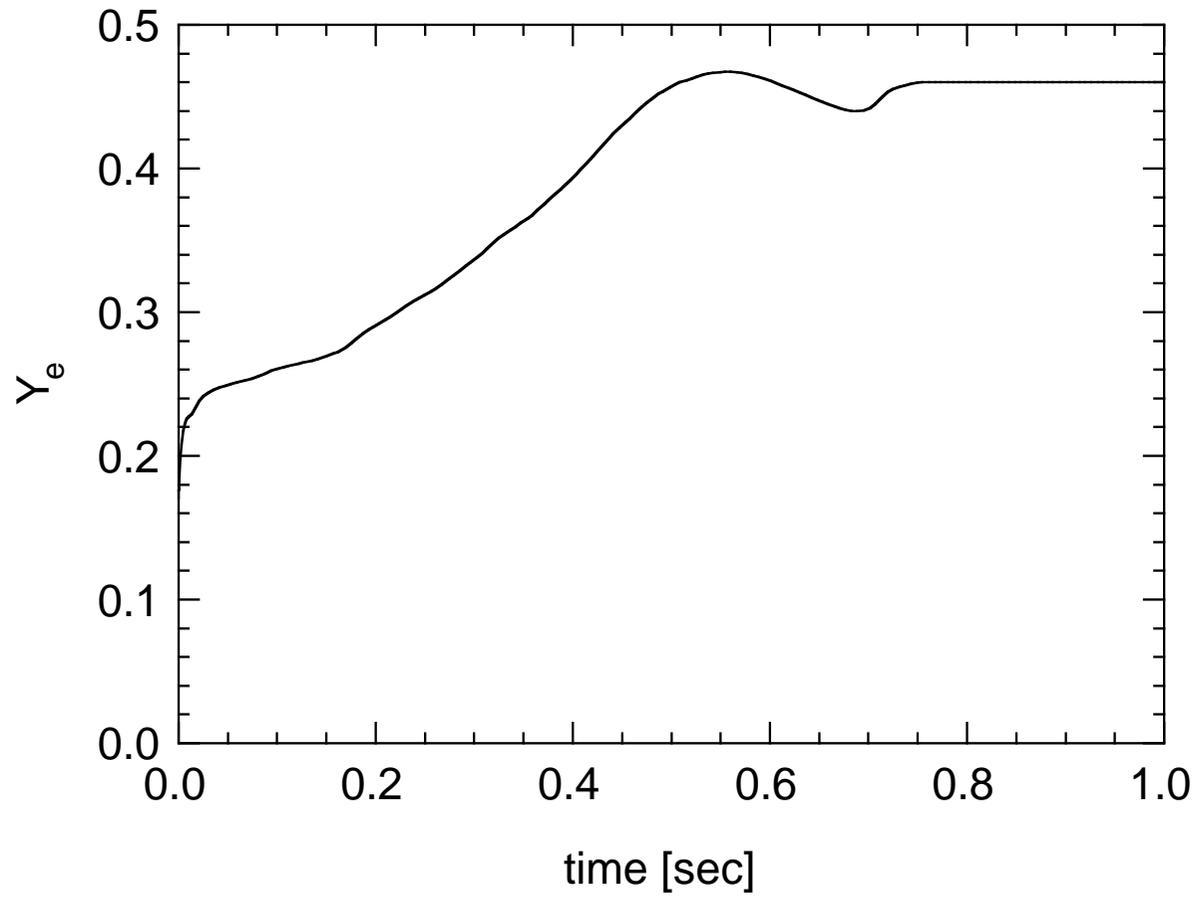
\item[Figure 4]
The electron fraction of a mass element for the same model 
as Fig. 1.  

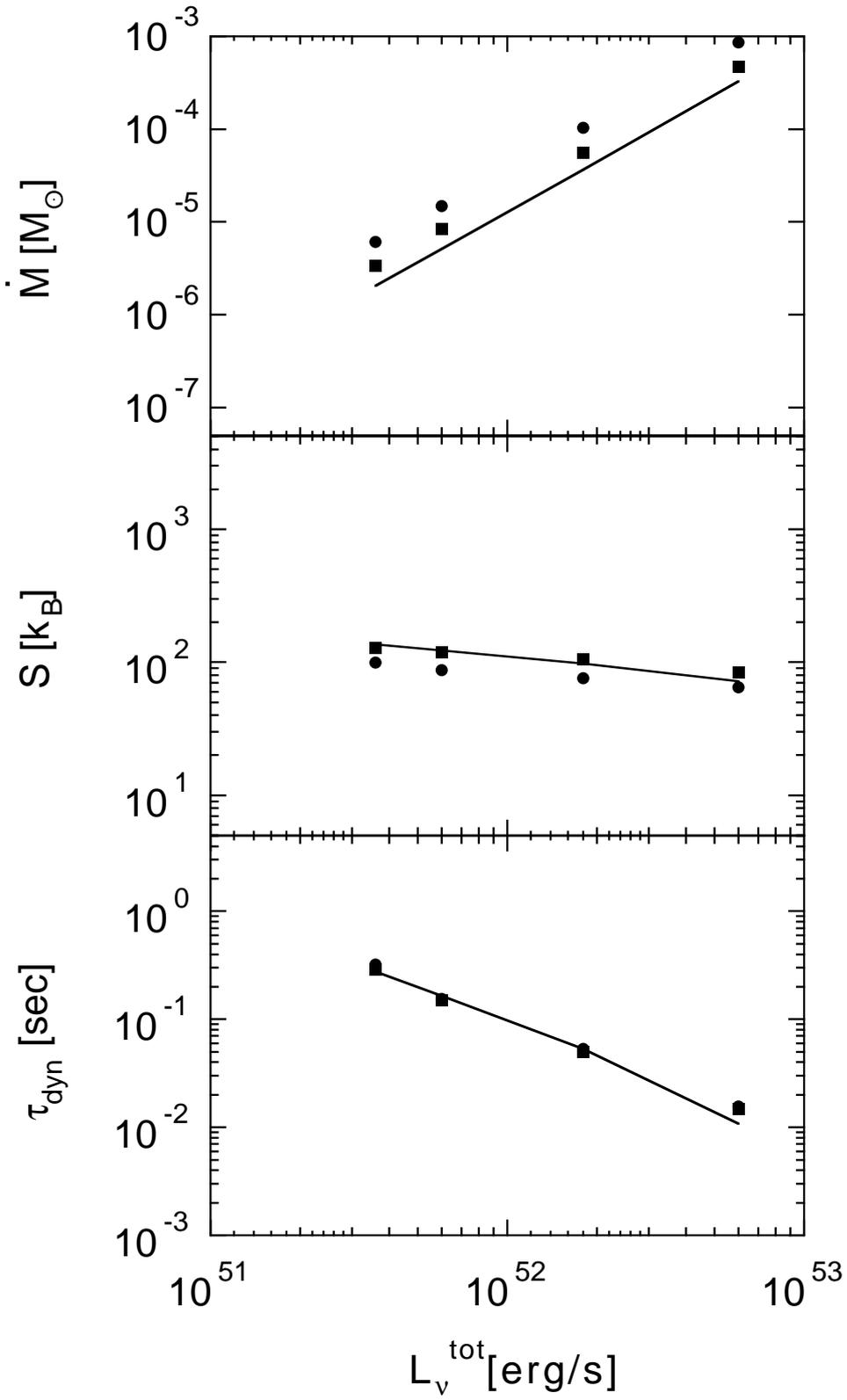
\item[Figure 5]
The luminosity 
dependence of the mass loss rate, entropy per baryon and 
expansion time scale for the models based on the proto-neutron 
star cooling simulations (symbols) 
and for the case of the analytic treatment (solid line).  
The solid circles are models with the baryon mass of 
$1.62 M_{\odot}$ and the solid squares with the baryon mass of 
$2.00 M_{\odot}$.  
The result of the analytic treatment is for the case with the 
gravitational mass of $1.4 M_{\odot}$ and the radius of 
10 km.  

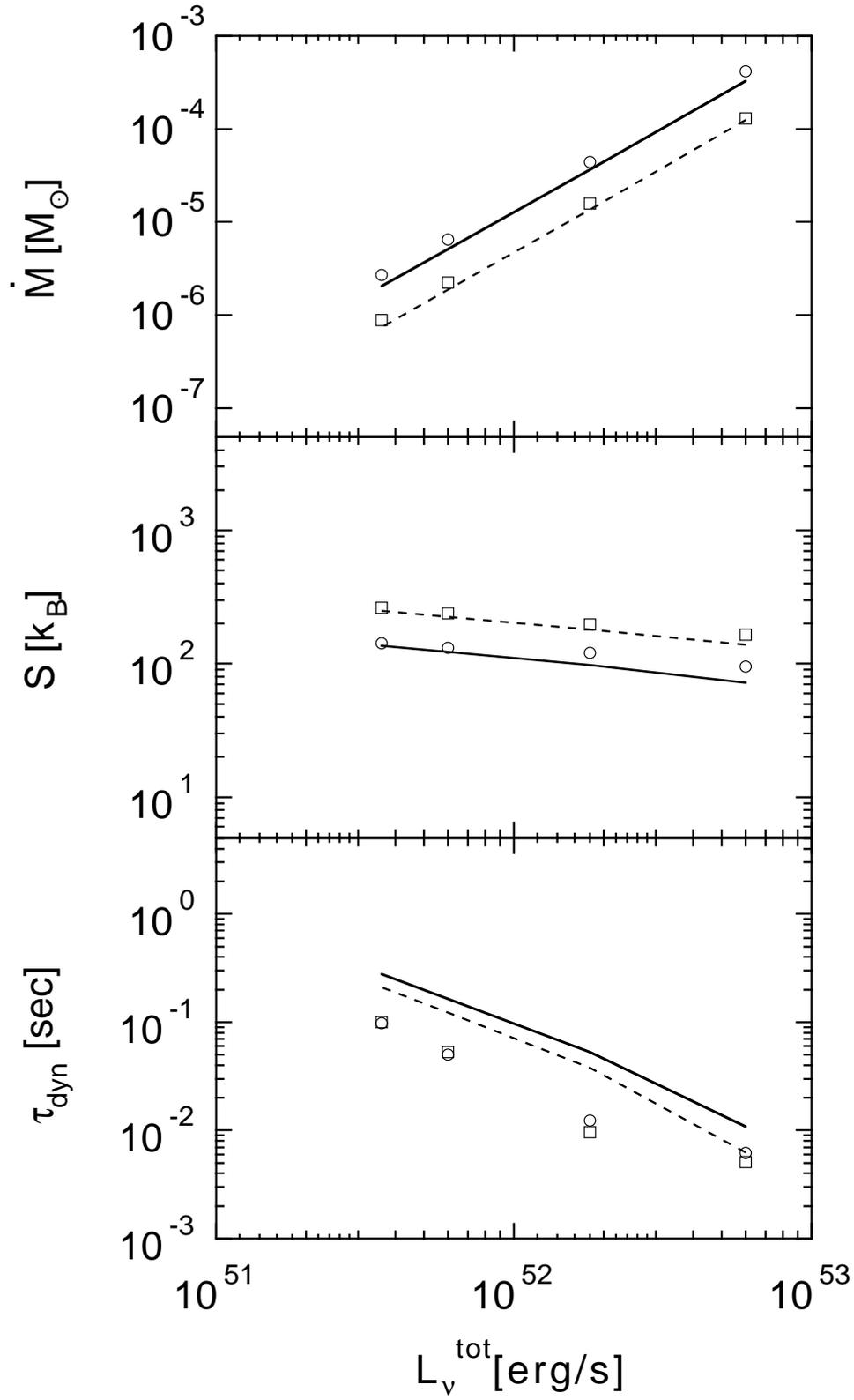
\item[Figure 6]
The luminosity 
dependence of the mass loss rate, entropy per baryon and 
expansion time scale for the models with given mass and radius 
(symbols) and for the analytic treatment (solid line).  
The open circles are models with the gravitational mass of 
$1.4 M_{\odot}$ and the open squares with the gravitational mass of 
$2.0 M_{\odot}$.  The solid and dashed lines show the results of 
the analytic treatment with the gravitational mass of 
$1.4 M_{\odot}$ and $2.0 M_{\odot}$, respectively.  
The radii are taken as 10 km in all cases.  

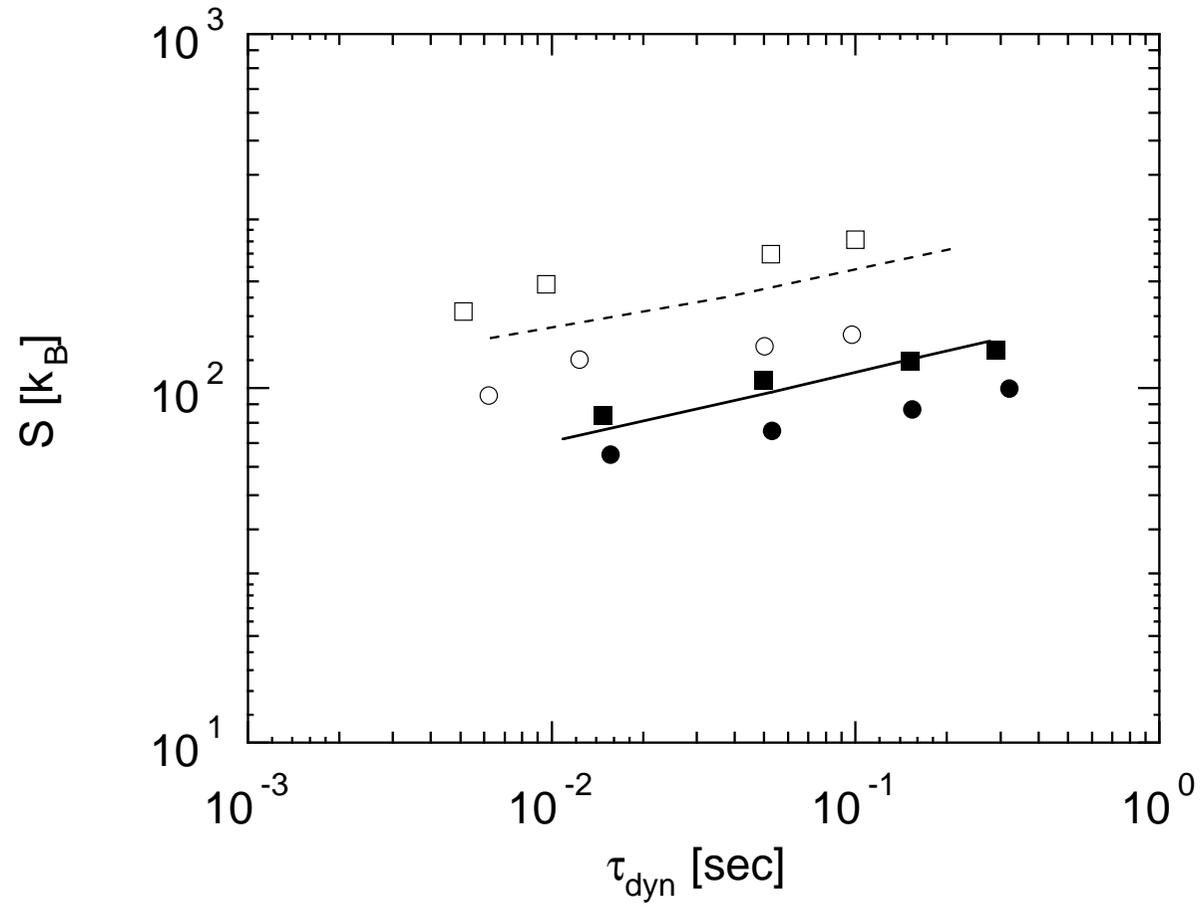
\item[Figure 7]
The expansion time scale ($\tau_{dyn}$) and entropy per baryon ($S$) 
at the temperature $T=0.5$ MeV.  
The solid circle and solid square symbols are for the models 
based on the proto-neutron star cooling simulations 
with the baryon mass of 
$1.62 M_{\odot}$ and $2.00 M_{\odot}$, respectively.  
The open circle and open square symbols are for the models 
with given mass and radius (the gravitational 
mass of $1.4 M_{\odot}$ and $2.0 M_{\odot}$, respectively).  
The solid and dashed lines show the results of the analytic treatment 
with the gravitational mass of $1.4 M_{\odot}$ and $2.0 M_{\odot}$, 
respectively.  
The radii are taken as 10 km in all cases.  

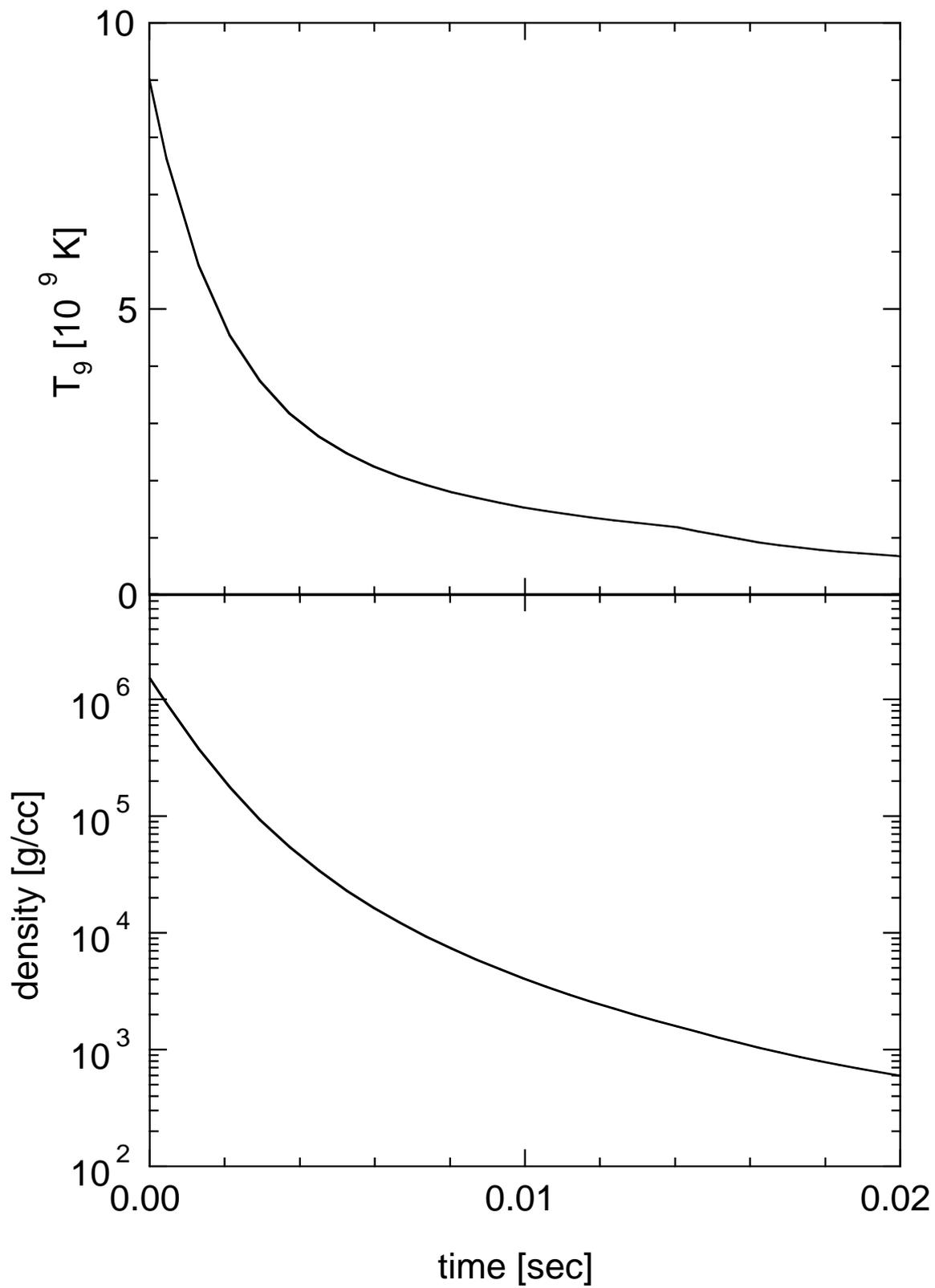
\item[Figure 8]
The time evolution of the temperature and density of a mass element 
taken from the model with the shortest expansion time scale 
(the gravitational mass of $2.0 M_{\odot}$, 
the radius of 10 km, and the total neutrino luminosity of 
$6 \times 10^{52}$ ergs/s).  
The time when the temperature becomes $T_{9}=9$ is defined to be 0.  

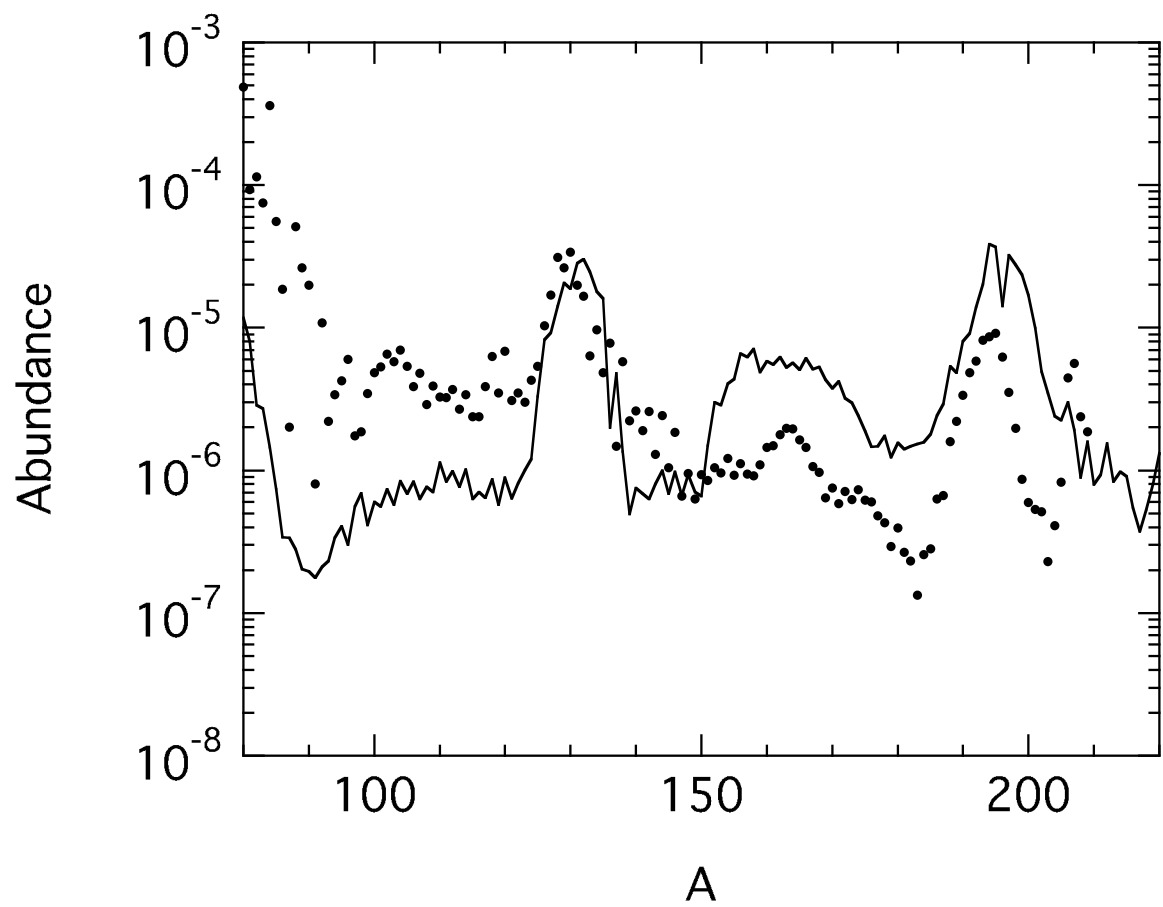
\item[Figure 9]
The abundance of the r-process elements obtained by the network 
calculation using the trajectory shown in Fig. 8.  
The scaled r-process abundances from K\"appeler, Beer and Wisshak (1989) 
are shown by dots for comparison.  

\end{description}

%%%%%%%%%%%%%%%%%%%%%%%%%%%%%%%%%%%%%%%%%%%%%%%%%%%%%%%%%%%%%%%%%%%%

\end{document}